\documentclass[mypaper,8pt,twoside]{CoAst}
\usepackage{epsf,graphicx,fancyhdr}
\renewcommand{\chaptermark}[1]{\markboth{#1}{}}

\newcommand{\kopf}{\small\itshape Comm. in Asteroseismology \\ Contribution to the Proceedings of the 38$^{th}$\,LIAC\,/\,HELAS-ESTA\,/\,BAG, 2008
}

\newcommand{\Authors}[1]{\begin{center}\normalsize\bf\sf #1 \end{center}}

\renewcommand{\author}[1]{\begin{center}\normalsize\bf\sf #1 \end{center}}
\newcommand{\Address}[1]{\begin{center}\small\sf #1 \end{center}}

\DeclareGraphicsExtensions{.eps,.jpg}

\newcommand{\Session}[1]{{\vspace{3mm}\small \noindent  \hspace*{3mm} Session: } #1 \normalsize}

	\newcommand{\five}{\small What can asteroseismology do to solve the problems }

\renewenvironment{abstract}{\section*{Abstract}\normalsize\sf}{}
\newcommand{\References}[1]{\begin{flushleft}{\large References\\}\vspace*{2mm}\small #1 \end{flushleft}}

\newcommand{\chapterCoAst}[2]{\chapter[\sf\normalsize #1\\ \footnotesize \hspace*{5mm}by #2 \sf\normalsize][]{#1\\}\rhead[\fancyplain{}{\sf\footnotesize \center{#1}}]{\fancyplain{}{\sffamily\thepage}}\lhead[\fancyplain{\kopf}{\sffamily\thepage}]{\fancyplain{\kopf}{\sf\footnotesize \center{#2}}}}

\renewcommand{\chaptername}{}

\newcommand{\acknowledgments}[1]{\vspace*{5mm}\noindent  \textbf{Acknowledgments.} #1}

\newcommand{\BV}{Brunt-V\"ais\"ala\ }
\newcommand{\simgt}{\lower.5ex\hbox{$\; \buildrel > \over \sim \;$}}

\newcommand{\msol}{\mbox{${\rm M}_{\odot}$}}
\newcommand{\msun}{\mbox{${\rm M}_{\odot}$}}

\def\rfr{\smallskip\par\noindent
        \hangindent=7truemm
        \hangafter=1}

\begin{document}
\sf

\chapterCoAst{Discriminating between overshooting and rotational mixing in massive stars: any help from asteroseismology?}%paper titel and page heading for even pages
{A. Miglio, J.\,Montalb\'an, P.\,Eggenberger, and A.\,Noels} %page heading for odd pages
\Authors{A.\,Miglio, J.\,Montalb\'an, P.\,Eggenberger and A.\,Noels}
\Address{Institut d'Astrophysique et de G\'eophysique \\
Universit\'e de Li\`{e}ge, All\'ee du 6 Ao\^{u}t 17 - B 4000 Li\`{e}ge - Belgique
}

\noindent
\begin{abstract}
Chemical turbulent mixing induced by rotation can affect the internal distribution of $\mu$ near the energy-generating core of main-sequence stars, having an effect on the evolutionary tracks similar to that of overshooting. However, this mixing also leads to a smoother chemical composition profile near the edge of the convective core, which is reflected in the behavior of the buoyancy frequency and, therefore, in the frequencies of gravity modes.
We show that for rotational velocities typical of main-sequence B-type pulsating stars, the signature of a rotationally induced mixing significantly perturbs the spectrum of gravity modes and mixed modes, and can be distinguished from that of overshooting.
The cases of high-order gravity modes in Slowly Pulsating B stars and of low-order g modes and mixed modes in $\beta$ Cephei stars are discussed.

\end{abstract}
\Session{ \five } % you can chose from, \one, \two, ... \six, \ESTA, \future or \poster
%\Objects{FG\,Vir, 44\,Tau, HD\,210111, ...}

\section{Introduction}
\label{sec:intro}
Asteroseismology of main-sequence B-type stars is now providing us with constraints on the structure and on the internal rotation rate of massive stars (see e.g. the recent review by \textit{Aerts~2008}). In particular, evidence for non-rigid rotation and for extra mixing near the core was found in several $\beta$ Cep pulsators (see e.g. \textit{Aerts et al.~2003}, \textit{Mazumdar\,et al.~2006}, \textit{Briquet et al.~2008}, \textit{Dziembowski \& Pamyatnykh 2008}). 
In this context, an even deeper insight into the internal structure of B stars could be obtained if the seismic probe provided by the observed oscillation modes is fine enough to differentiate the near-core extra mixing due to overshoot from that induced by rotation.

\begin{figure}
\begin{center}
\includegraphics[width=0.45\textwidth]{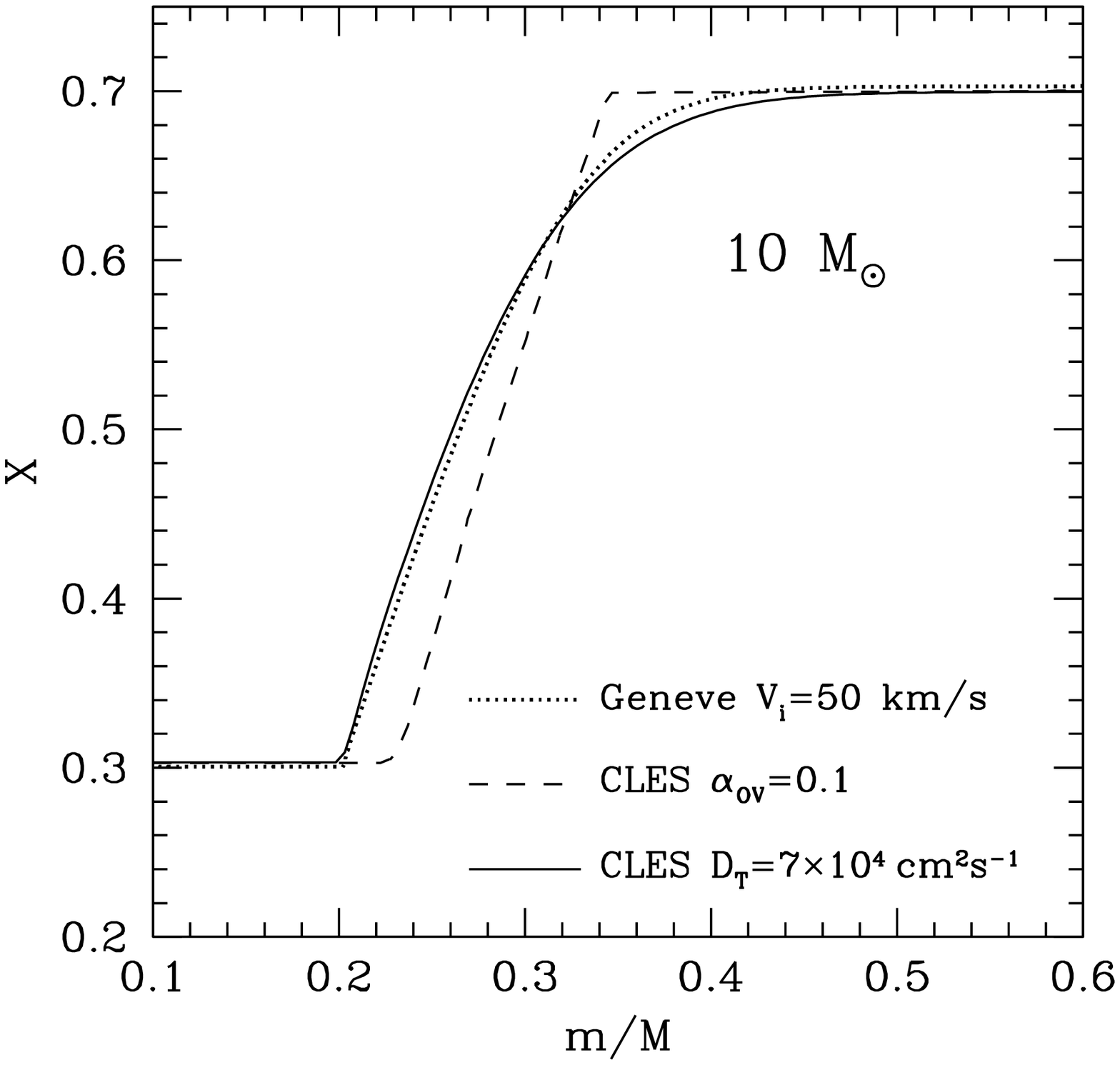}
\includegraphics[width=0.45\textwidth]{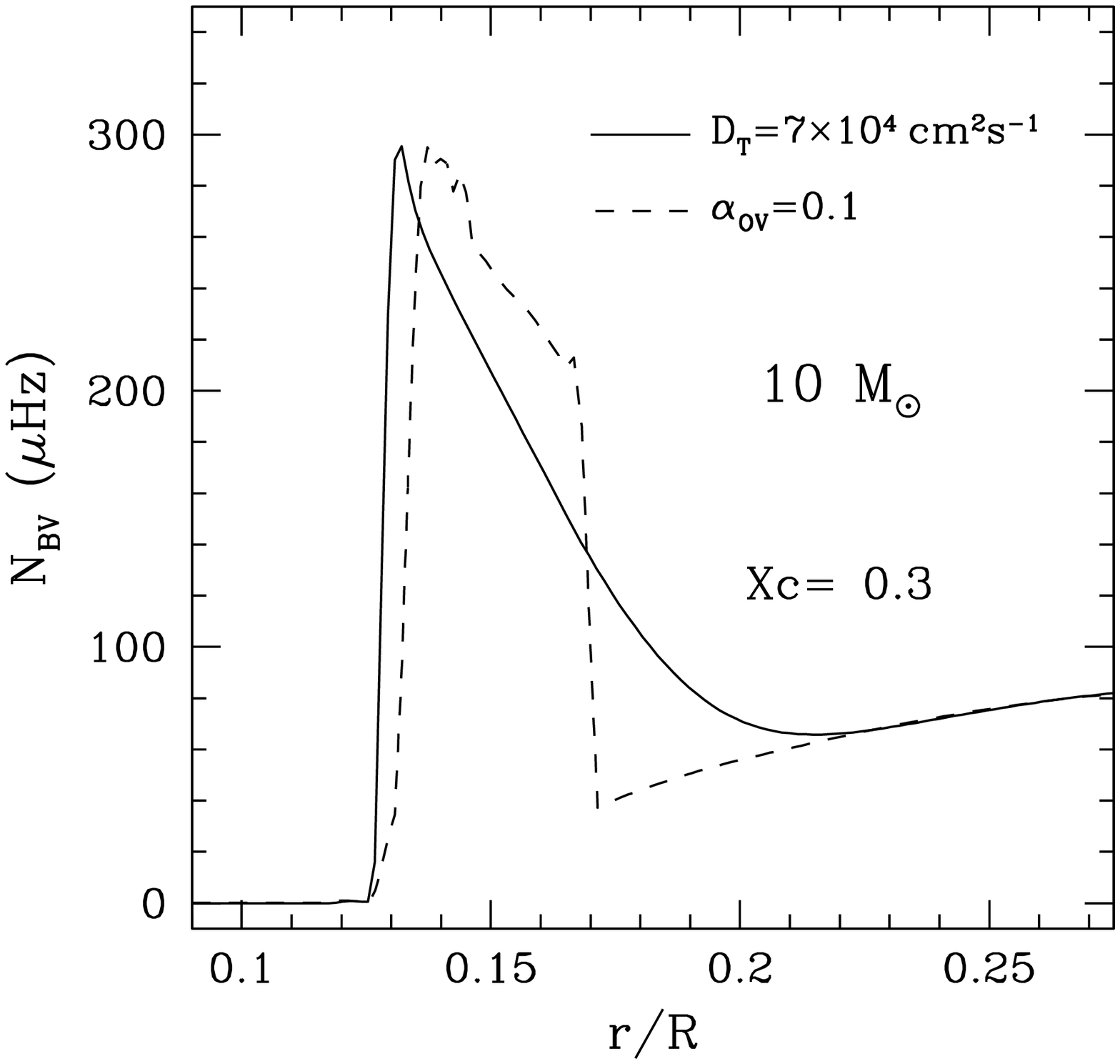}
\caption{Left panel: Hydrogen mass fraction profile in the central regions of 10~\msol\ models with $X_{\rm C}\simeq0.3$. The different lines correspond to models computed with {\sc cles}, with overshooting (dashed line), and turbulent diffusion coefficient $D_{\rm T}$ (solid line), and with the Geneva code with an initial rotational velocity of 50~km~s$^{-1}$ (dotted line). Right panel: \BV\ frequency profile in the central regions of the 10~\msol\ {\sc cles} models shown in the left panel.}
\label{fig:xprofil}
\end{center}
\end{figure}

With the aim of studying the effects of different extra-mixing processes on the frequency spectra of SPB and  $\beta$ Cep pulsators we focus, respectively, on models of 6 and 10~\msol.
We consider the effect of overshooting and of turbulent mixing near the core that could be induced by the effects of rotation on the transport of chemicals.
The models with overshooting were computed with {\sc cles} (Code Li\'egeois d'Evolution Stellaire, \textit{Scuflaire et al.~2008}). In this code the thickness of the overshooting layer $\Lambda_{\rm OV}$ is parameterized in terms of the local pressure scale height $H_p$: $\Lambda_{\rm OV}=\alpha_{\rm OV}\times(\min(r_{\rm cc},H_p(r_{\rm cc}))$, where $r_{\rm cc}$ is the radius of the convective core given by the Schwarzschild criterion and $\alpha_{\rm OV}$ is a free parameter. In the overshooting region mixing is assumed to be instantaneous and the temperature stratification radiative.

As a first approach, the turbulent mixing has been modelled in {\sc cles} by a diffusion process with a parametric turbulent diffusion coefficient $D_{\rm T}$ that is uniform inside the star and independent of age. We computed also models using the Geneva code that includes the treatment of rotation and related transport processes as described in \textit{Eggenberger et al.~2008}. The comparison between the two series of models shows that the chemical composition profiles in the central regions provided by a uniform and time independent diffusion coefficient represent a first approximation, at least for massive stars,  of the effect that would be produced by such a rotationally induced  chemical transport (see Fig. \ref{fig:xprofil}).
The values of the parameter $D_{\rm T}$ were chosen in order to be close to the value of the chemical diffusion coefficient near the core  provided by the Geneva models (see \textit{Montalb\'an et al. 2008} for more details on this parametrization).

SPBs and $\beta$~Cep pulsators are considered as slow or moderate rotators. SPBs show a typical rotational velocity  of 25~km~s$^{-1}$ (\textit{Briquet et al.~2007}), whereas the range of projected rotational velocity in $\beta$~Cep stars extends from  0 to 300~km~s$^{-1}$ with an average of 100~km~s$^{-1}$ (\textit{Stankov~\& Handler 2005}).   The Geneva  code calculations for 10~\msol\ models provide values of the chemical diffusion coefficient near the convective core with  $X_{\rm C}=0.3$,  of the order of   $5\times10^4$~cm$^2$s$^{-1}$  for a rotational velocity on the zero-age main sequence (hereafter named initial rotational velocity, $V_i$) of 20~km~s$^{-1}$, $7\times10^4$~cm$^2$s$^{-1}$ for $V_{i}$=50~km~s$^{-1}$, and $1.6\times10^5$~cm$^2$s$^{-1}$ for $V_i$=100~km~s$^{-1}$. On the other hand, the effect of an initial rotational velocity of 25km~s$^{-1}$ on the central hydrogen distribution of  a 6~\msol\ model is well mimicked by a $D_{\rm T} \sim 5000$~cm$^2$s$^{-1}$. The results presented in this paper concern mainly the parametric models with $D_{\rm T}=5\times10^3$~cm$^2$s$^{-1}$ for SPB models and $D_{\rm T}=7\times10^4$~cm$^2$s$^{-1}$ for  $\beta$~Cep ones, and they will be compared with overshooting models closely located in the HR diagram, that means $\alpha_{\rm OV}$=0 for the SPB case and 0.1 for the $\beta$~Cep one.

As shown in Fig. \ref{fig:xprofil}, models located at the same position in the HR diagram, but computed with overshooting or with turbulent mixing, show a significantly different chemical composition profile near the core (see Fig.~\ref{fig:xprofil} left panel) and consequently a different behaviour of the sharp feature in the \BV\ frequency ($N$) located in the $\mu$-gradient region (Fig.~\ref{fig:xprofil} right panel).

\section{High-order g modes in Slowly Pulsating B stars}
As presented in \textit{Miglio et al.~(2008a)} (hereafter Paper I), the periods of high-order gravity modes in main-sequence stars can be related, by means of analytical expressions, to the detailed characteristics of the $\mu$-gradient region that develops near the energy-generating core, and thus to the mixing processes that affect the behaviour of $\mu$ in the central regions.
The period spacing ($\Delta P$) of high-order g modes can be described as a superposition of a constant term predicted by the first order approximation by \textit{Tassoul~(1980)} and periodic components directly related to the location and sharpness of $\nabla_\mu$.

We recall that in the asymptotic approximation presented in \textit{Tassoul (1980)} the periods of high-order gravity modes are given by $P_{kl}=\pi^2\Pi_0\,L^{-1}(2k+cte)$, where $L=[\ell(\ell+1)]^{1/2}$ (with $\ell$ the mode degree), $k$ the order of the mode and $\Pi_0^{-1}=\int_{x_0}^1{\frac{N}{x'}dx'}$, where $x'$ is the normalized radius and $x_0$ corresponds to the boundary of the convective core.

\label{sec:high}
\begin{figure}
\begin{center}
\resizebox{0.32\hsize}{!}{\includegraphics[angle=0]{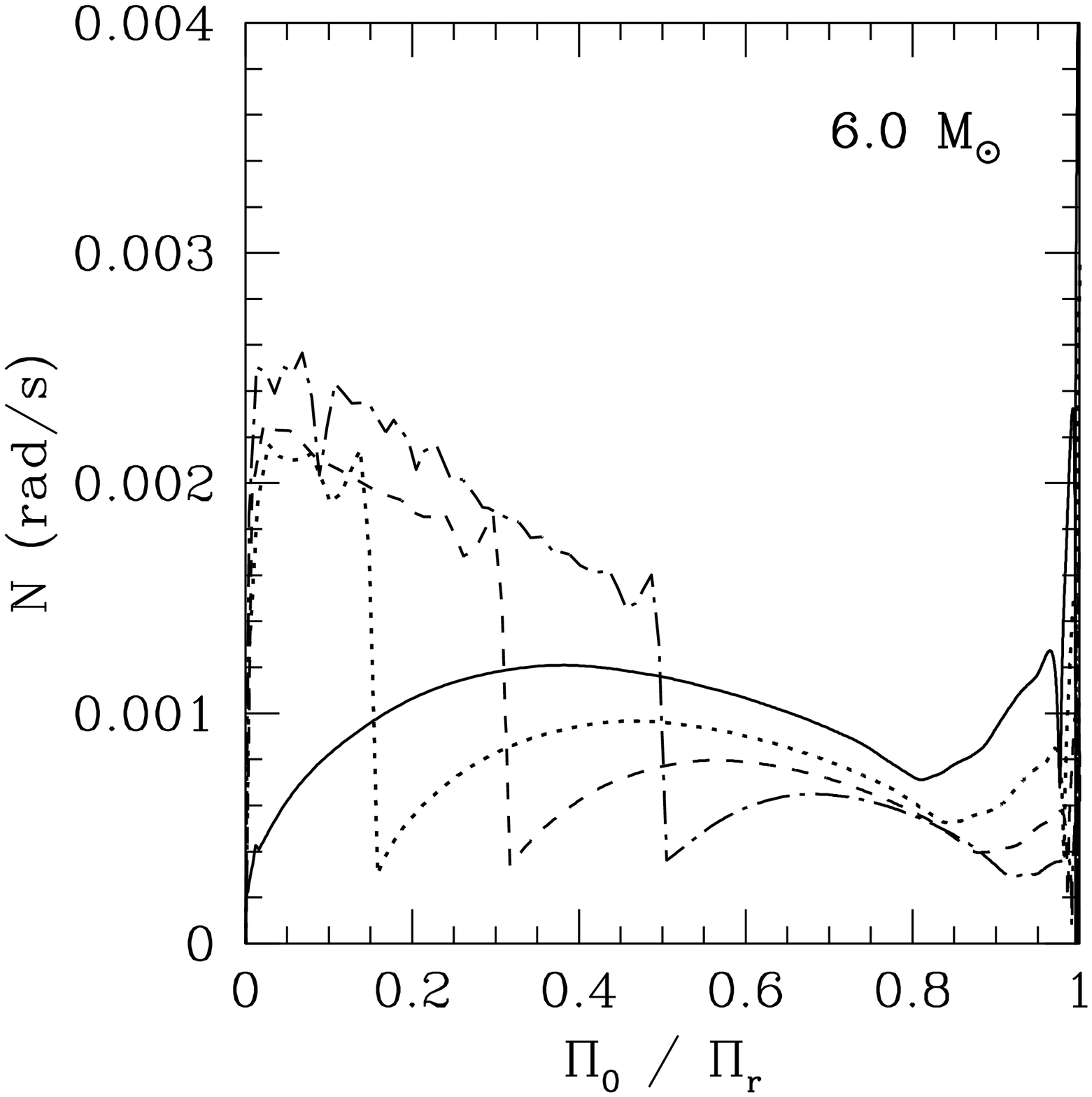}}
\resizebox{0.32\hsize}{!}{\includegraphics[angle=0]{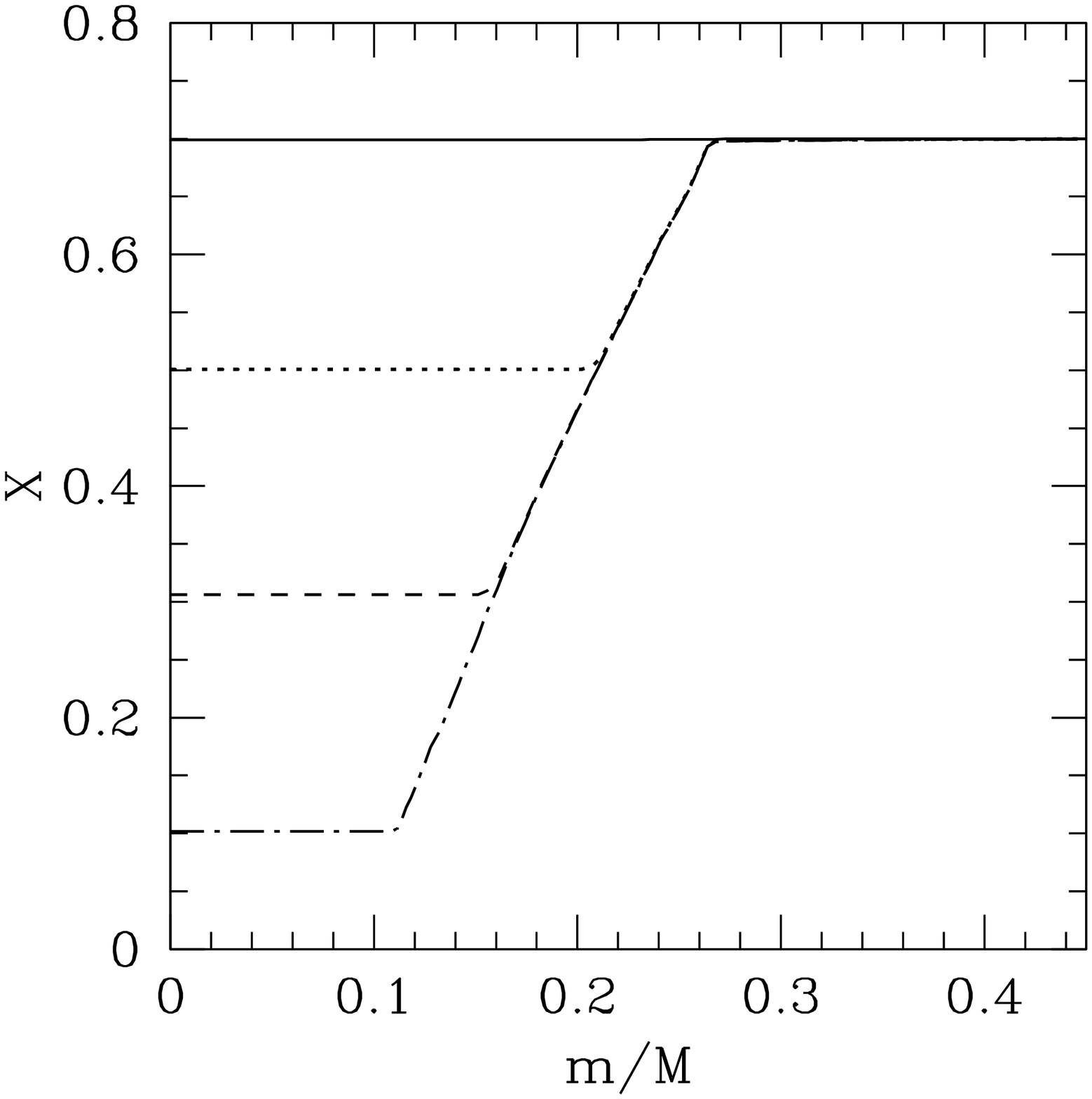}}
\resizebox{0.32\hsize}{!}{\includegraphics[angle=0]{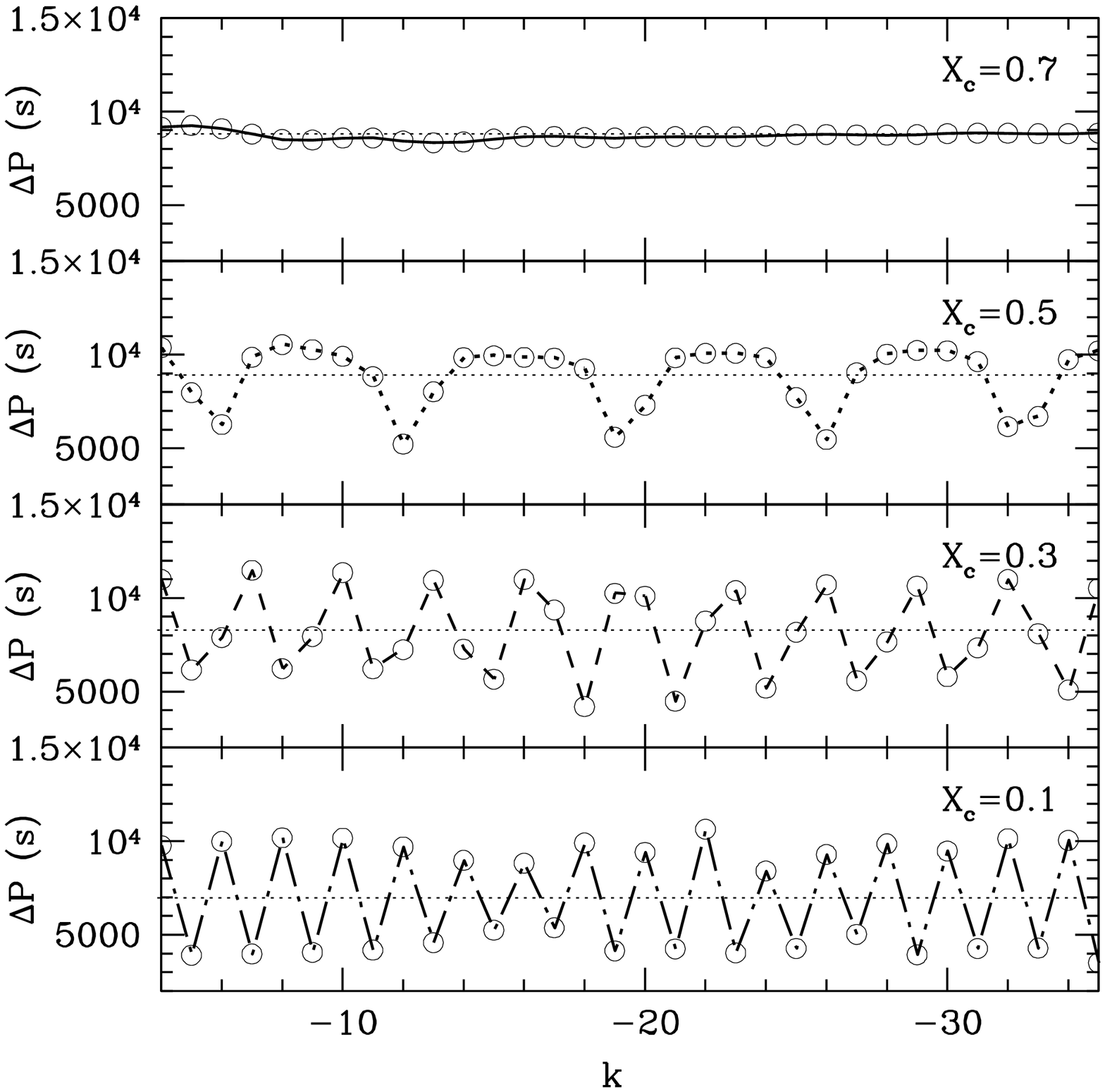}}
\resizebox{0.32\hsize}{!}{\includegraphics[angle=0]{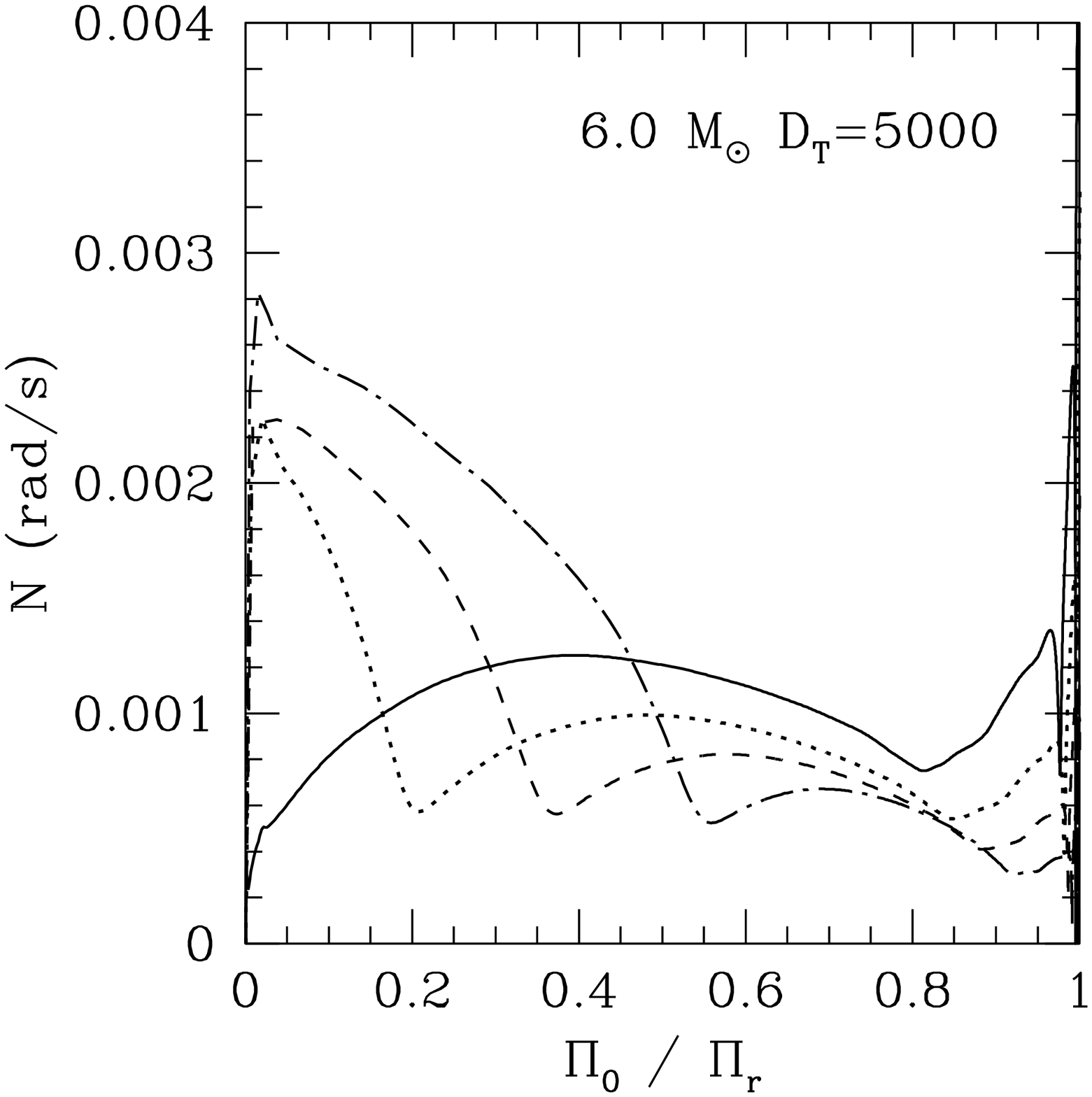}}
\resizebox{0.32\hsize}{!}{\includegraphics[angle=0]{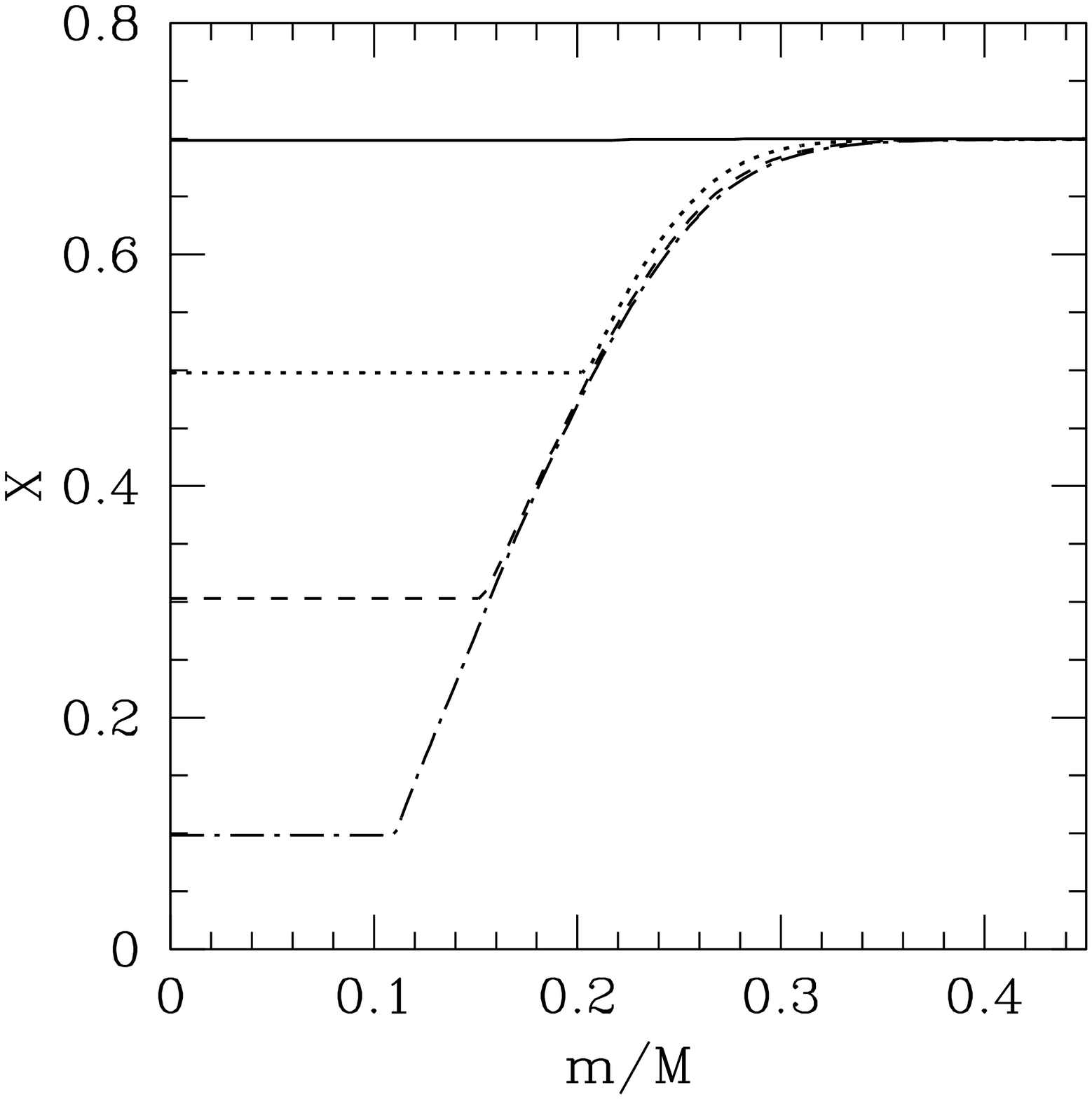}}
\resizebox{0.32\hsize}{!}{\includegraphics[angle=0]{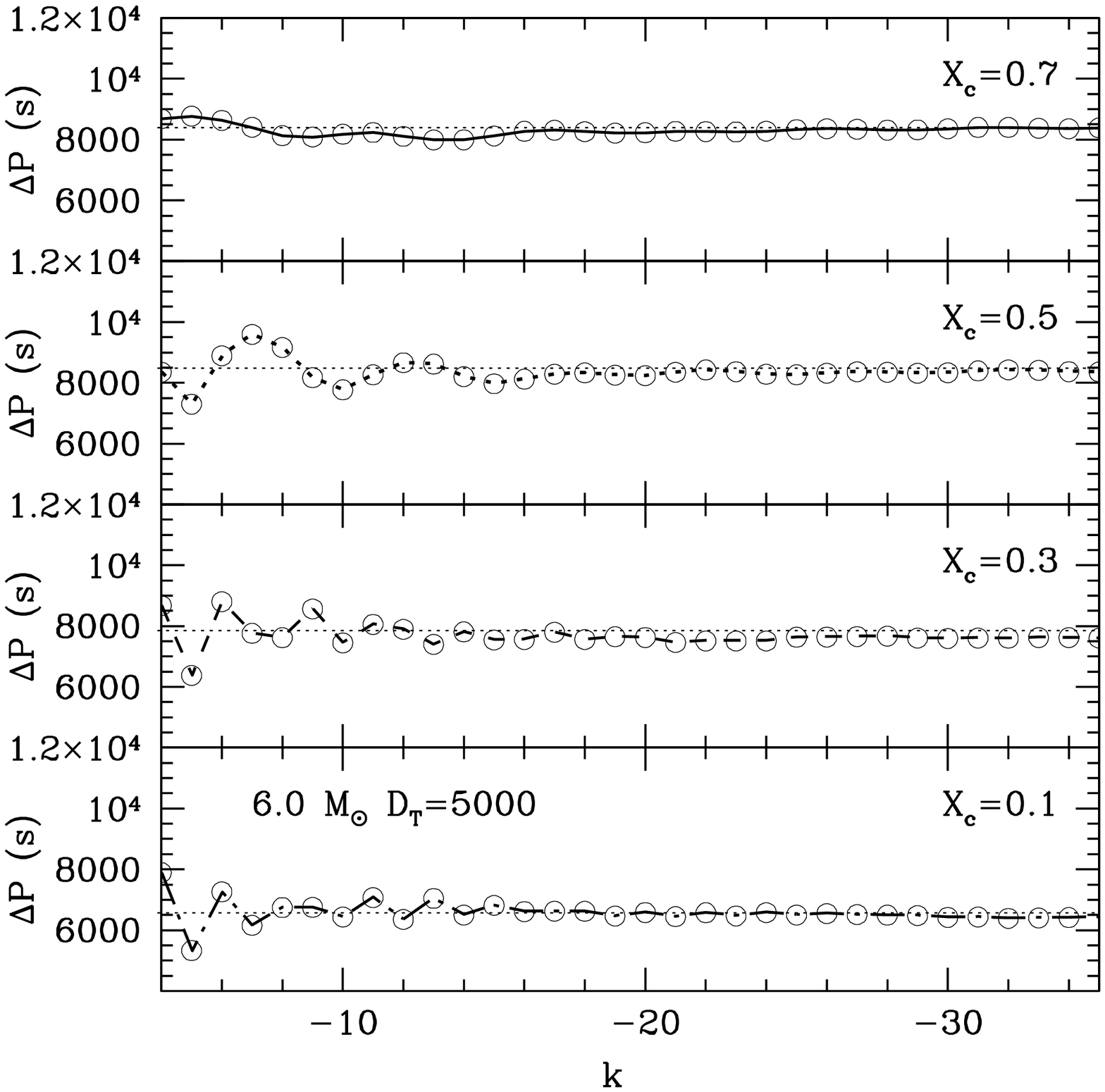}}

\caption{\small Behaviour of the \BV frequency (left panels), of the hydrogen abundance profile (middle panels) and of the $\ell=1$ g-mode period spacing in models of 6 \msun (right panels). We consider several models along the main sequence with decreasing central hydrogen abundance ($X_{\rm c}$= 0.7, 0.5, 0.3 and 0.1). In the right panels the uppermost model has $X_c=0.7$ and the lowermost $X_c= 0.1$. Upper panels refer to models computed with no extra mixing, whereas lower panels in models of 6.0 \msol computed with a turbulent diffusion coefficient $\rm D_{\rm T}=5000$ cm$^2$ s$^{-1}$.}
\label{fig:6}
 \end{center}
\end{figure}
In main-sequence stars $\nabla\mu$ depends on the size of the convective core and on extra-mixing processes (e.g. overshooting, diffusion, turbulent mixing) that may alter the chemical composition profile in the central region of the star.
We here briefly describe the effect of rotationally induced mixing by considering models of a 6 \msol\ star computed including turbulent diffusion near the core (described by a turbulent-diffusion coefficient $\rm D_{\rm T}=5\times10^3$ cm$^2$ s$^{-1}$), which mimics the effects of an initial rotational velocity of 25km~s$^{-1}$.
In Fig. \ref{fig:6} we compare the period spacing of such models (lower-right panel) with models that share the same location in the HR diagram, but are computed without any extra-mixing (upper-right panel). We notice that turbulent mixing has a substantial effect on the period spacing: the amplitude of the periodic components in $\Delta P$ becomes a decreasing function of the radial order $k$. This behaviour can be easily explained by the analytical approximation in Paper I, provided that the smoother $\mu$ profile (and thus $N$, see Fig. \ref{fig:6}) is taken into account in the analysis.

\section{Low-order g modes and mixed modes in $\beta$ Cephei stars}
\label{sec:low}
\begin{figure*}
\begin{center}
\includegraphics[width=0.8\textwidth]{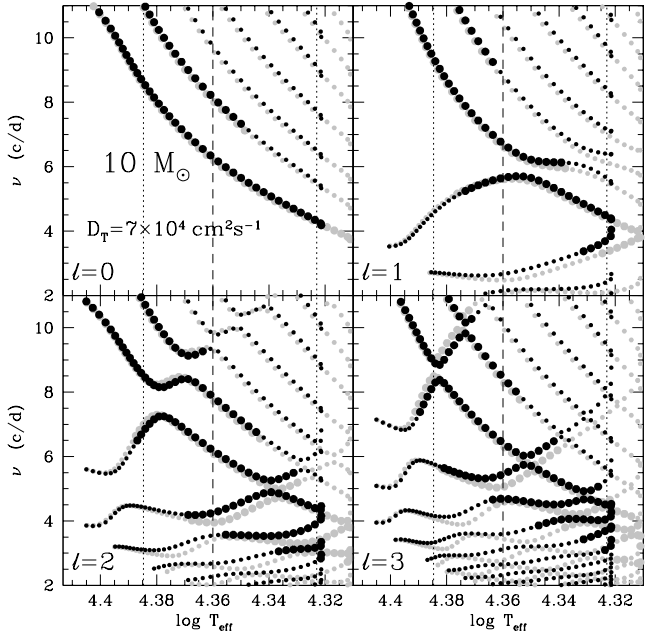}

\caption{Frequencies of  pulsation modes with angular degree $\ell=0-3$ as a function of $\log T_{\rm eff}$ for main-sequence models of a 10~\msol\ star. Gray dots correspond to the frequencies of models computed with an overshooting parameter $\alpha_{\rm OV}=0.1$, whereas black dots are the frequencies of models computed with a turbulent diffusion coefficient $D_{\rm T}=7\times10^4$ cm~s$^{-2}$. Excited modes are represented by thicker symbols. The vertical lines indicate the effective temperature of models with a hydrogen  mass fraction at the center  of the order of  0.5, 0.3 and 0.1 (left to right).}
\label{fig:exciD7}
\end{center}
\end{figure*}

\begin{figure}
 \parbox{0.5\textwidth}{
\begin{center}
\includegraphics[width=0.45\textwidth]{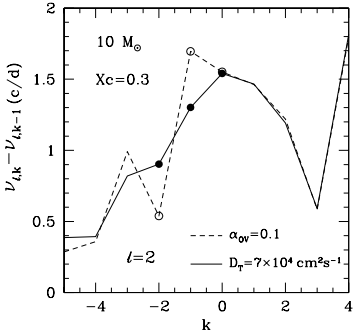}
\end{center}
}
 \parbox{0.45\textwidth}{
\caption{Frequency difference (in c/d) between modes with consecutive radial order ($k$) and degree $\ell=2$ for the 10~\msol\ {\sc cles} models in Fig.~\ref{fig:xprofil}.}
\label{fig:dnu}
}
\end{figure}
As recalled in the previous section, changes of $\mu$ induced by the turbulent mixing acting near the core lead to significant variations in the properties of the spectrum of high-order g modes.  Although the approximation presented in Paper I and used to relate the sharpness of the ``bump'' in the \BV\ frequency to the properties of g modes is no longer valid for low-order modes, the analytical description of gravity modes mentioned there is still able to qualitatively relate some properties of low-order g-mode spectra to the characteristics of the $\mu$-gradient region near the core (see \textit{Miglio et al. 2008b}).

During the main sequence, the combined action of nuclear reactions and convective mixing leads to a chemical composition gradient at the boundary of the convective core, to a decrease of $\Pi_0$, and thus to an increase of the frequencies of gravity modes. The latter interact with pressure modes of similar frequency and affect the properties of non-radial oscillations by the so-called avoided-crossing phenomenon. The  modes undergoing an avoided crossing (mixed modes) are therefore sensitive probes of the core structure of the star.

As presented in \textit{Montalb\'an et al.~(2008)}, we describe the effects of extra-mixing on the oscillation spectrum of a typical $\beta$~Cep star by comparing the properties of low-order g and p modes in models computed with overshooting and with chemical turbulent diffusion. Since the parameters $\alpha_{\rm OV}$ and $D_{\rm T}$ were chosen to lead to similar evolutionary tracks on the HR diagram, such comparison allows us to remove the differences in the frequencies due to a different stellar radius. In Fig.~\ref{fig:exciD7} we plot the oscillation frequencies  for 10~\msol\ models along the main-sequence phase.
As expected, the differences  between frequencies of overshooting models and turbulent mixing ones is very small for pressure modes, while significant differences appear for gravity and mixed modes. These differences increase with the radial order of g modes and, for a given frequency, with the angular degree of the modes. In Fig. \ref{fig:exciD7} we have also marked with thicker symbols the frequencies of  modes that the non-adiabatic oscillation code {\sc MAD} (\textit{Dupret et al.~2003}) predicts to  be excited. Here we show only the case with $D_{\rm T}=7\times10^4$~cm$^2$s$^{-1}$ but computations with lower or higher efficiency of the turbulent mixing show that the differences between overshooting and turbulent-mixing models increase with the value of $D_{\rm T}$.

What is most relevant from an asteroseismic point of view, is the change in the distance between consecutive frequencies of g-modes or modes of mixed p-g character ($\Delta\nu$). In Fig.~\ref{fig:dnu} we  plot these differences for the $\ell=2$ modes of the 10~\msol\ models with an effective temperature $T_{\rm eff}\simeq 22550$~K ($X_{\rm C}\simeq 0.3$). The dots indicate differences computed between pairs of excited modes. Therefore the difference of $\Delta \nu$ that we can expect between models with sharp $\nabla_\mu$ (overshooting models for instance) and models with a chemical composition gradient smoothed by the effect of, for instance, a slow rotation ($V_{\rm rot}\sim 50$~km/s) is of the order of 0.4~c/d ($\sim 5\,\mu$Hz), much larger than the precision of present and forthcoming  observations.

\section{Conclusions}
We presented how the frequencies of gravity modes and mixed modes in $\beta$ Cep and SPB stars depend on the detailed characteristics of the $\mu$-gradient region that develops near the energy generating core, and thus on the mixing processes that can affect the behaviour of $\mu$ in the central regions. In particular we have shown that for rotational velocities typical of main-sequence B-type pulsating stars, the signature of a rotationally induced mixing significantly perturbs the spectrum of gravity modes and mixed modes, and can be distinguished from that of overshooting.

Such a sensitivity of g-mode oscillation frequencies to near-core mixing can provide an additional constraint to the modelling of massive stars with rotation, especially when coupled with seismic inferences on the internal rotational profile (see e.g. the review by \textit{Goupil~\& Talon 2009}), and with constraints on chemical enrichments in the photosphere (e.g. \textit{Maeder et al.~2008} and \textit{Morel 2008}).
Further investigations are however needed to assess under which observational conditions such information can be recovered from the oscillation frequencies, given realistic uncertainties on stellar global parameters, errors on oscillation frequencies and, in the case of SPBs, given the severe influence of rotation on the spectrum of g modes.

\acknowledgments{A.M. and J.M. acknowledge financial support from the Prodex-ESA Contract Prodex 8 COROT (C90199). A.M. is a \emph{Charg\'e de Recherches} of the FRS-FNRS.
P.E. is thankful to the Swiss National Science Foundation for support.
}

\References{
\rfr Aerts, C. 2008, IAUS, 250, 237
\rfr Aerts, C., Thoul, A., Daszynska, J. et al. 2003, Science, 300, 1926
\rfr Briquet, M., Hubrig, S., De Cat, P. et al. 2007, A\&A, 466, 269
\rfr Briquet, M., Morel, T., Thoul, A. et al. 2007, MNRAS, 381, 1428
\rfr Dupret, M.-A., De Ridder, J., De Cat, P., et al. 2003, A\&A, 398, 677
\rfr Dziembowski, W., Pamyatnykh, A. 2008, MNRAS, 385, 206
\rfr Eggenberger, P., Meynet, G., Maeder, A, et al. 2008, Ap\&SS, 316, 43
\rfr Goupil, M.J., Talon., S., 2008, CoAst, 158, in press
\rfr Maeder, A.,  Meynet, G., Ekstrom, S., \& Georgy, C. 2008, CoAst 158, in press (arXiv:0810.0657)
\rfr Mazumdar, A., Briquet, M., Desmet, M., \& Aerts, C. 2006, A\&A, 459, 589
\rfr Miglio, A., Montalb\'an, J., Noels, A., \& Eggenberger, P. 2008a, MNRAS, 386, 1487
\rfr Miglio, A., Montalb\'an, J., Eggenberger, P., \& Noels, A. 2008b, AN, 329, 529
\rfr Montalb\'an, J., Miglio, A.,  Eggenberger, P., \& Noels, A. 2008, AN, 329, 535
\rfr Morel, T. 2008, CoAst, 158, in press (arXiv:0811.4114)
\rfr Scuflaire, R., Montalb\'an, J., Th\'eado, S., et al. 2008, ApSS, 316, 149
\rfr Stankov, A., \& Handler, G. 2005, ApJS, 158, 193
\rfr Tassoul, M. 1980, ApJS, 43, 469
}

\end{document}